\documentclass[%
 reprint,
 amsmath,
    amssymb,
 aps,
prx,
]{revtex4-2}

\usepackage{dcolumn}
\usepackage{bm}

\usepackage{graphicx}
\usepackage[utf8]{inputenc}
\usepackage[T1]{fontenc}
\usepackage{siunitx}
\usepackage{physics}
\usepackage{hyperref}
\usepackage{cleveref}
\usepackage{lineno}

\newcommand{\idEst}{\textit{i.e.}}
\newcommand{\ExempliGratia}{\textit{e.g.}}
\newcommand{\via}{\textit{via}}
\newcommand{\inSilico}{\textit{in silico}}
\newcommand{\cilia}{\textit{cilia}}
\newcommand{\flagella}{\textit{flagella}}
\newcommand{\PAeruginosa}{\textit{P. aeruginosa}}
\newcommand{\EColi}{\textit{E. coli}}
\newcommand{\VAlgo}{\textit{V. alginolyticus}}
\newcommand{\ESPResSo}{\textit{ESPResSo}}
\newcommand{\porespy}{\textit{porespy}}

\newcommand{\rntlong}{run-and-tumble}
\newcommand{\RnTlong}{Run-and-tumble}

\newcommand{\RnRlong}{Run-and-reverse}
\newcommand{\rnrlong}{run-and-reverse}

\newcommand{\RRFlong}{Run-reverse-flick}
\newcommand{\rrflong}{run-reverse-flick}

\newcommand{\rwnlong}{reverse-when-stuck}
\newcommand{\Rwnlong}{Reverse-when-stuck}

\newcommand{\partclPos}{\vb{r}}
\newcommand{\frictionTrans}{\gamma_\text{t}}
\newcommand{\vSwim}{v_\text{swim}}
\newcommand{\partclDirec}{\vu{u}}
\newcommand{\kB}{k_\text{B}}
\newcommand{\temperature}{T}
\newcommand{\kbT}{\kB\temperature}
\newcommand{\noiseTrans}{\boldsymbol{\eta}^\text{t}}

\newcommand{\partclAngVel}{\vb*{\omega}}
\newcommand{\frictionRot}{\gamma_\text{r}}
\newcommand{\activeAngVel}{\omega_\text{act}}
\newcommand{\orthoVec}{\vu{n}}
\newcommand{\noiseRot}{\boldsymbol{\eta}^\text{r}}
\newcommand{\diffCoeffRot}{D_\text{rot}}
\newcommand{\rotationTimescale}{\tau_\text{rot}}

\newcommand{\bactNBeads}{N_\text{beads}}
\newcommand{\bactLBody}{l_\text{body}}
\newcommand{\bactRBead}{r_\text{body}}
\newcommand{\dynVisc}{\mu}

\newcommand{\timeStep}{\delta t}
\newcommand{\timeSlice}{\Delta t}

\newcommand{\tRun}{t_\text{run}}
\newcommand{\avgTRun}{\expval{\tRun}}
\newcommand{\tTumble}{t_\text{tumble}}
\newcommand{\avgTTumble}{\expval{\tTumble}}
\newcommand{\tumbleDRot}{D_\text{rot, tumble}}
\newcommand{\tumbleAngle}{\Theta_\text{tumble}}
\newcommand{\actualTumbleAngle}{\tumbleAngle^*}
\newcommand{\flickAngle}{\Theta_\text{flick}}
\newcommand{\tFlick}{t_\text{flick}}
\newcommand{\tMemory}{t_\text{memory}}

\newcommand{\radiusHydroSphere}{R_\text{sphere}}
\newcommand{\nBacts}{N}
\newcommand{\nEnsemble}{N_\text{ensemble}}
\newcommand{\simDuration}{\mathcal{T}}
\newcommand{\localThickness}{\tau}
\newcommand{\porosity}{\phi}

\newcommand{\diffCoeffEff}{D_\text{eff}}
\newcommand{\meanPoreRadius}{r_\text{p}}
\newcommand{\MSD}{\ensuremath{\text{MSD}}}

\newsavebox{\dotbox}
\newcommand{\outerdot}[1]{%
  \sbox\dotbox{$#1$}\dot{\usebox\dotbox}}

\usepackage[normalem]{ulem}
\usepackage{xcolor}
\newcommand{\revOneScratch}[1]{}
\newcommand{\revOneAdd}[1]{#1}
\newcommand{\revTwoScratch}[1]{}
\newcommand{\revTwoAdd}[1]{#1}

\newcommand{\revOneScratchRoundTwo}[1]{}
\newcommand{\revOneAddRoundTwo}[1]{#1}
\newcommand{\revTwoScratchRoundTwo}[1]{}
\newcommand{\revTwoAddRoundTwo}[1]{#1}

\begin{document}
    \title{Optimal motility strategies for self-propelled agents to explore porous media}
    \author{Christoph Lohrmann}
    \email{clohrmann@icp.uni-stuttgart.de}
    \affiliation{Institute for Computational Physics, University of Stuttgart, 70569 Stuttgart, Germany}%
    \author{Christian Holm}
    \email{holm@icp.uni-stuttgart.de}
    \affiliation{Institute for Computational Physics, University of Stuttgart, 70569 Stuttgart, Germany}%

    \date{\today}

    \begin{abstract}
        Micro-robots for, \textit{e.g.}, biomedical applications, need to be equipped with motility strategies that enable them to navigate through complex environments.
        Inspired by biological microorganisms we recreate motility patterns such as \rnrlong, \rntlong{} or \rrflong{} applied to active rod-like particles \inSilico.
        We investigate their capability to efficiently explore disordered porous environments with various porosities and mean pore sizes ranging down to the scale of the active particle.
        By calculating the effective diffusivity for the different patterns, we can predict the optimal one for each porous sample geometry.
        We find that providing the agent with \revTwoScratch{the ability to sense position for a certain time and to make a decision based on its observation}
        \revTwoAdd{very basic sensing and decision making capabilities} yields a motility pattern outperforming the biologically inspired patterns for all investigated porous samples.
    \end{abstract}

    \maketitle

    \section{Introduction}
    \label{sec:introduction}
    Evolution has equipped microorganisms with a variety of motility patterns that allow them to explore their environment efficiently for various tasks.
    For example, bacteria live in soil or larger host organisms where they search their environment for nutrients and surfaces to colonize.
    A very common environmental constraint is confinement, both in the habitat of biological microswimmers and in the application domain of their artificial counterparts, \textit{e.g.}, micro-robots.
    Bacteria are used for engineering applications in porous media such as crack sealing, soil stabilisation and contamination remediation~\cite{choi17a, phillips16a,mujah17a, priyadarshanee21a}.
    It is envisioned that artificial micro-robots or micro-swimmers can in the future act as micro-surgeons and perform medical tasks inside human tissue~\cite{nelson10a, patra13a}.
    In each case, the bacteria or micro-robots, from now on called agents, first have to traverse a highly confining, disordered porous environment before they can fulfill their function.

    Self-propulsion is a necessary ingredient for the efficient exploration of such an environment, however, self-steering can improve the performance significantly.
    Microorganisms can achieve directional control by changing the beating patterns and synchronisation of their propelling \cilia~\cite{josef06a} or \flagella~\cite{schwarz16a, gong20a}.
    Many basic artificial microswimmers are unable to steer, especially if their propulsion mechanism relies on chemical reactions~\cite{paxton04a, howse07a, li14b}.
    However, progress has been made in the control of individual artificial swimmers that are actuated by light~\cite{lozano16a, baeuerle20a, massana22a} or magnetic fields~\cite{cheang17a, fernandez20a, keicheang14a}, endowing them with a steering feature.

    Biological microswimmers are known to possess various motility patterns~\cite{johansen02a, stocker12a}, \idEst, strategies to use a combination of self-propulsion and self-steering to navigate through their environment:
    \PAeruginosa{} and many marine bacteria can reverse their locomotion and perform a \rnrlong{} pattern in which they alternate between forward and backward swimming~\cite{johansen02a}.
    Bacterium \EColi{} interrupts its forward swimming mode ("run") with reorientation events ("tumble"), where the bacterium rotates before continuing to swim in a new direction~\cite{berg72a}.
    \VAlgo{} alternates between swimming forward, swimming backward and flicking its orientation by \SI{90}{\degree}, a pattern called \rrflong~\cite{xie11a}.
    In the following we will use these motility patterns as a starting point to investigate optimal strategies for porous media exploration and navigation.

    The spreading behaviour of active particles with different motility patterns has been well studied in unconfined fluids~\cite{taktikos13a} and weakly confined environments~\cite{bechinger16a}.
    Diffusive properties under strong confinement have also been the subject of a number of experimental and theoretical works:
    Zeitz \textit{et al.} investigated in detail the mean-squared displacement of disk-like active Brownian particles (straight swimmers subject to rotational diffusion) in a porous environment close to the percolation threshold~\cite{zeitz17a}.
    Bhattacharjee \& Datta tracked \EColi{} cells in three-dimensional porous media and found that the bacterial trajectories cannot be identified as \rntlong{} anymore, but they rather found a sequence of hopping events through the channels, with the bacteria being intermittently trapped in small pores~\cite{bhattacharjee19a,bhattacharjee19b}.
    Theoretical studies of \rntlong{}-swimmers in porous media find a maximal effective diffusivity by optimizing the duration of runs for specific pore configurations~\cite{reichhardt14a, licata16a,bertrand18a, irani22a}.
    Similarly, numerical simulations of \rnrlong{}-swimmers show that the optimal run length can be inferred from the distribution of the lengths of straight paths in a porous medium~\cite{kurzthaler21b}.

    While the aforementioned works have optimized the parameters of specific patterns for porous media exploration, we will attempt here to optimize the motility pattern itself.
    We study the qualitative features of different patterns when used by otherwise identical agents in various three-dimensional, disordered environments.
    We cover the range of all relevant pore sizes from bulk fluid to confinement ranging down to the size of the micro-swimmer.
    Using the insights gained from our analysis of biologically inspired motility patterns, we develop a new pattern which requires the agents to be capable of sensing \revTwoScratch{their environment and making an intelligent decision} \revTwoAdd{whether they are trapped or not}.
    This pattern, which can be deployed by artificial autonomous self-propelled agents, performs best across all investigated environments, and can be a basis for developing further optimal navigation strategies.

    \section{Results}
    \label{sec:results}

    \subsection{Agent model and motility patterns}
    \begin{figure}
        \includegraphics[width=\linewidth]{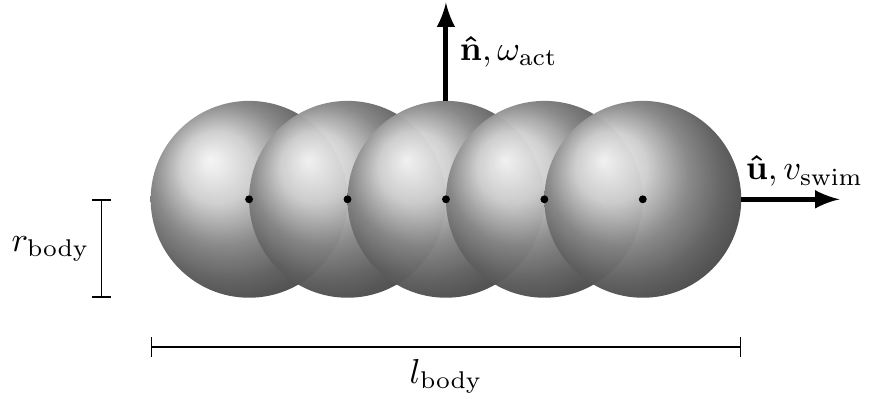}
        \caption{Schematic representation of the geometry of the rigid self-propelled agent.}
        \label{fig:model}
    \end{figure}
\begin{figure}
        \includegraphics[width=\linewidth]{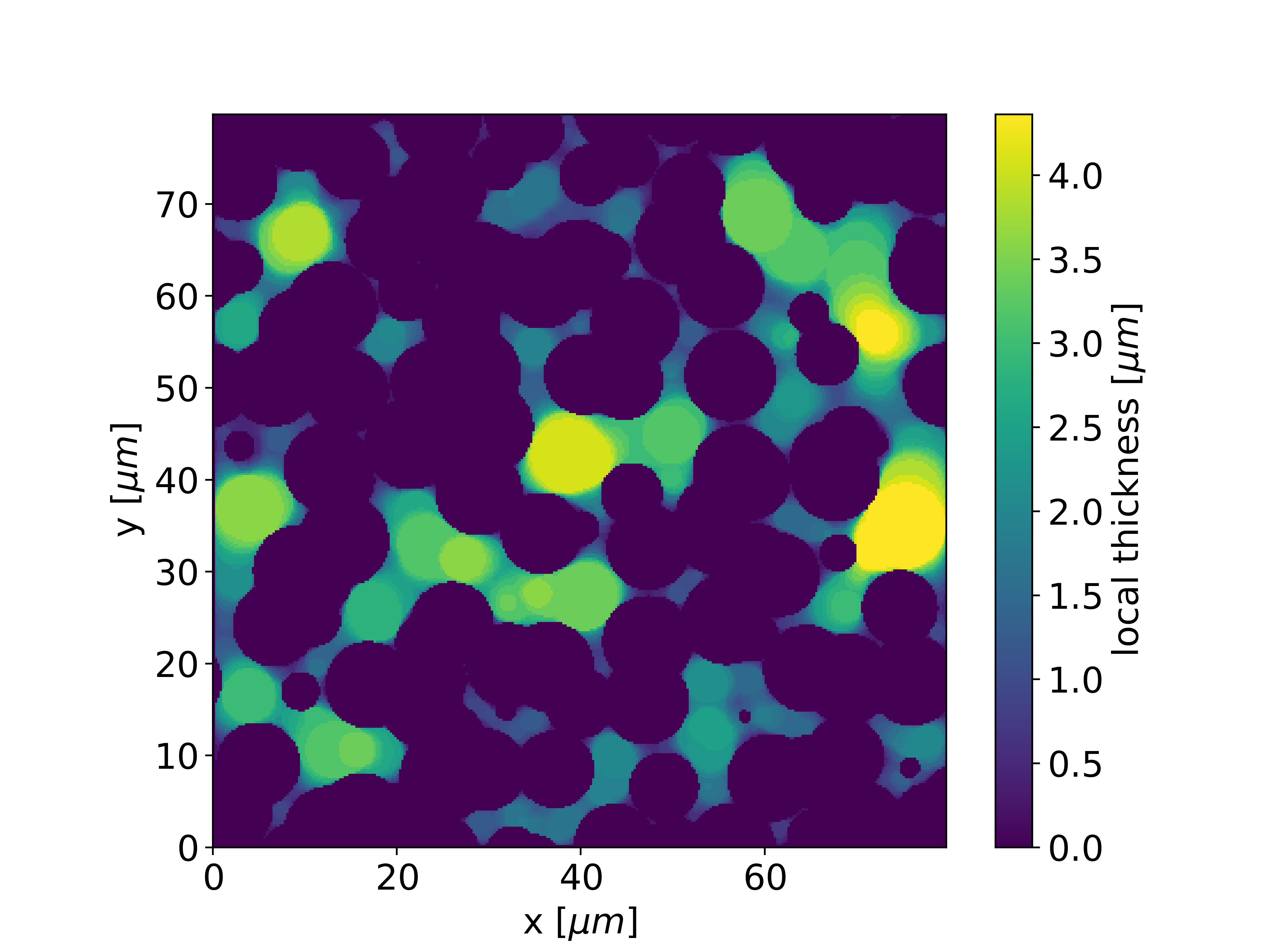}
        \caption{A two-dimensional slice through a typical randomly generated porous geometry with mean pore radius $\meanPoreRadius \approx \SI{2.6}{\micro\meter}$. The colors indicate the local thickness, \revOneScratch{ see color bar.} \revOneAdd{a measure for the pore size (see Methods).}}
        \label{fig:porous_slice_local_thickness}
    \end{figure}

    We simulate $\nBacts=100$ individual active agents in three dimensions by modeling them as rod-like, rigid-body particles with length $\bactLBody = \SI{2}{\micro\meter}$ and radius $\bactRBead= \SI[parse-numbers = false]{1/3}{\micro\meter}$, as shown in \cref{fig:model}.
    The agents perform translational and rotational Brownian motion and are subject to repulsive interactions with their porous environment.
    Additionally, we apply time-dependent active forces and torques to achieve self-propulsion with speed $\vSwim$ and self-steering with \revTwoScratch{rotation rate} \revTwoAdd{angular velocity} $\activeAngVel$ (see Methods for a detailed description of the equations of motion).
    The porous environment is modelled by randomly placing overlapping spherical obstacles with radius $\radiusHydroSphere = \revOneScratch{10} \revOneAdd{15} \bactRBead$ within the simulation domain.
    An example is shown in \cref{fig:porous_slice_local_thickness}\revTwoAdd{.
    For a detailled description of our porous media model, the control parameters and the derived quantities} \revTwoScratch{the interaction potential between agents and obstacles}, see Methods.

\begin{figure*}
        \includegraphics[width=\linewidth]{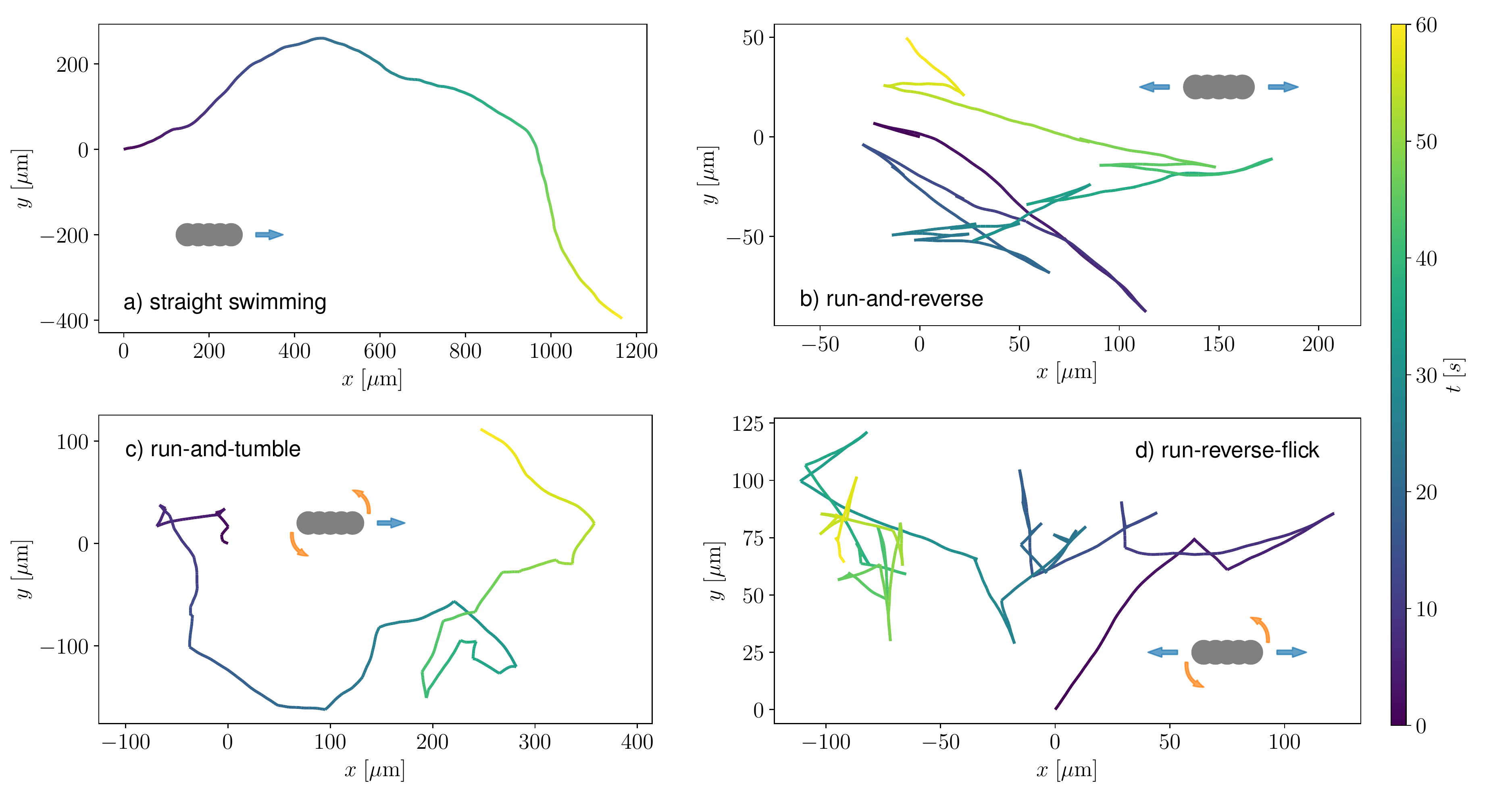}
        \caption{Two-dimensional projections of example trajectories without confinement for the four biologically inspired motility patterns.
        The pictograms (not to scale) show the phases involved in the respective pattern, \idEst, forward swimming, backward swimming and rotation.
        The trajectory for \rwnlong{} is not depicted, as this pattern reduces to straight swimming if there are no pores to get trapped in.
        For easier distinction, the temperature is reduced by a factor of 60 compared to the simulations used in our analysis.}
        \label{fig:example_trajs}
    \end{figure*}

    We create motility patterns by combining phases of self-propulsion and self-rotation, prescribing the durations of the phases and their temporal sequence.
    In the following, we list the algorithms of the patterns used in this study, example trajectories are shown in \cref{fig:example_trajs}.

    \paragraph{Straight swimming}
    A constant force along the symmetry axis is applied, there is no active rotation.
    The only source of randomness in the trajectory is the translational and rotational diffusion.
    Aside from the anisotropic shape of the self-propelled agent, this pattern is an implementation of a 3D active Brownian particle (ABP).

    \paragraph{\RnRlong}
    With this motility pattern, agents swim at constant speed $\vSwim$ ("run") but can reverse their swimming direction, realised by a change in sign of the self-propulsion velocity $\vSwim \to - \vSwim$.
    The reversal algorithm thus implies that agents reverse their swimming propulsion and not the direction of their body, as observed in nature~\cite{hintsche17a, taylor74a}, and so no active torques are applied.
    The durations $\tRun$ of runs are commonly found to be exponentially distributed for bacteria~\cite{duffy97a, berg93a}, therefore we draw them from a distribution
    \begin{equation}
        p(\tRun) = \frac{1}{\avgTRun} \exp(-\tRun/ \avgTRun).
        \label{eq:run_time_exp_distr}
    \end{equation}
    Here, $\avgTRun$ is the average run duration, which is the only adjustable parameter of the \rnrlong{} pattern.

    \paragraph{\RnTlong}
    The \rntlong{} pattern allows agents to swim straight ("run") and actively change their orientation ("tumble") at distinct times, a strategy employed by, \ExempliGratia, \EColi~\cite{berg72a}.
    Our numerical algorithm follows Lee \textit{et al.}~\cite{lee19a}, we only give a brief summary here:
    The durations $\tRun, \tTumble$ of runs and tumbles are exponentially distributed and drawn from distributions analogous to \cref{eq:run_time_exp_distr}.
    They are characterised by their respective means $\avgTRun$ and $\avgTTumble$.
    The model for tumbling is based on the assumption that during a tumble the rods perform rotational Brownian motion with an increased rotational diffusion coefficient $\tumbleDRot$.
    With this assumption, a distribution of tumble angles $\tumbleAngle$ can be calculated analytically for each tumble duration.
    The tumble rotational diffusion coefficient together with the average tumble duration are the control parameters that 
    determine the average angle of reorientation \revTwoScratch{$\expval{\tumbleAngle}$.} \revTwoAdd{according to~\cite{lee19a}
        \begin{equation}
                \expval{\cos(\tumbleAngle)} = \frac{1}{2 \tumbleDRot \avgTTumble+ 1}.
            \end{equation}}
    In our implementation, a tumble duration is drawn, and the associated tumble angle distribution is calculated.
    Then, a tumble angle is drawn from this distribution and an active torque with $\activeAngVel = \tumbleAngle/\tTumble$ is applied to the rod such that, in the absence of thermal noise and obstacles, the correct angle of rotation is achieved within the tumble duration.
    The direction $\orthoVec$ of the active torque is orthogonal to the particle orientation $\partclDirec$, and its azimuthal angle in the particle frame of reference is drawn at random from $[0, 2\pi)$.

    \paragraph{\RRFlong}
    This motility pattern can be found in marine bacteria \VAlgo{}~\cite{xie11a} and combines elements from \rnrlong{} and \rntlong.
    Here, runs (of exponentially distributed durations $\tRun$) are interrupted by both, reversals and flicks.
    A flick is a tumble with a constant angle $\flickAngle = \pi/2$ and duration $\tFlick$.
    Reversals and flicks occur in alternating fashion.

    \paragraph{\Rwnlong}
    Leaving the realm of motility patterns that occur in nature, we propose a hypothetical optimal pattern for porous media navigation that combines straight swimming and reversals.
    For this pattern, the agent must be endowed with sensing capabilities, a way to store a memory over a limited amount of time, and an intelligence unit to make simple decisions.
    Together, these capabilities enable smart reactions to the environment beyond following a predetermined order of self-propulsion and -rotation.
    Using a position sensor, the agent constructs a memory of its trajectory within a time frame $\tMemory$.
    If it did not move more than one rod length $\bactLBody$ in that time, a reversal is triggered.
    Upon reversal, the memory is reset.

    This algorithm is used as a representative of the whole class of motility patterns in which the agent is able to sense if it is stuck in a pore.
    A position sensor is not necessarily required, agents could also obtain this information from a sensor for swimming speed.
    Mechanical sensors on the agent body or propulsion mechanism such as the ones found in bacteria~\cite{gordon19a} could determine a trapped state as well.

    \subsection{Effective diffusivity} \label{subsec:effective-diffusivity}
    From the scale of the different trajectories in \cref{fig:example_trajs} one can already get a qualitative understanding of how efficient agents can explore unconfined spaces depending on the strategy they employ.
    To quantify the efficiency of exploration in both, unconfined space and porous media, we calculate the mean-squared displacement (\MSD)
    \begin{equation}
        \MSD(t) = \frac{1}{\nBacts} \sum_{i=1}^{\nBacts} \frac{1}{\simDuration-t} \int_{0}^{\simDuration-t} \abs{\partclPos_i(t'+t)-\partclPos_i(t')}^2 \dd{t'},
    \end{equation}
    where $\simDuration$ is the duration of the simulation, and $\partclPos_i$ the center of mass position of agent $i$.

\begin{figure}
        \includegraphics[width=\linewidth]{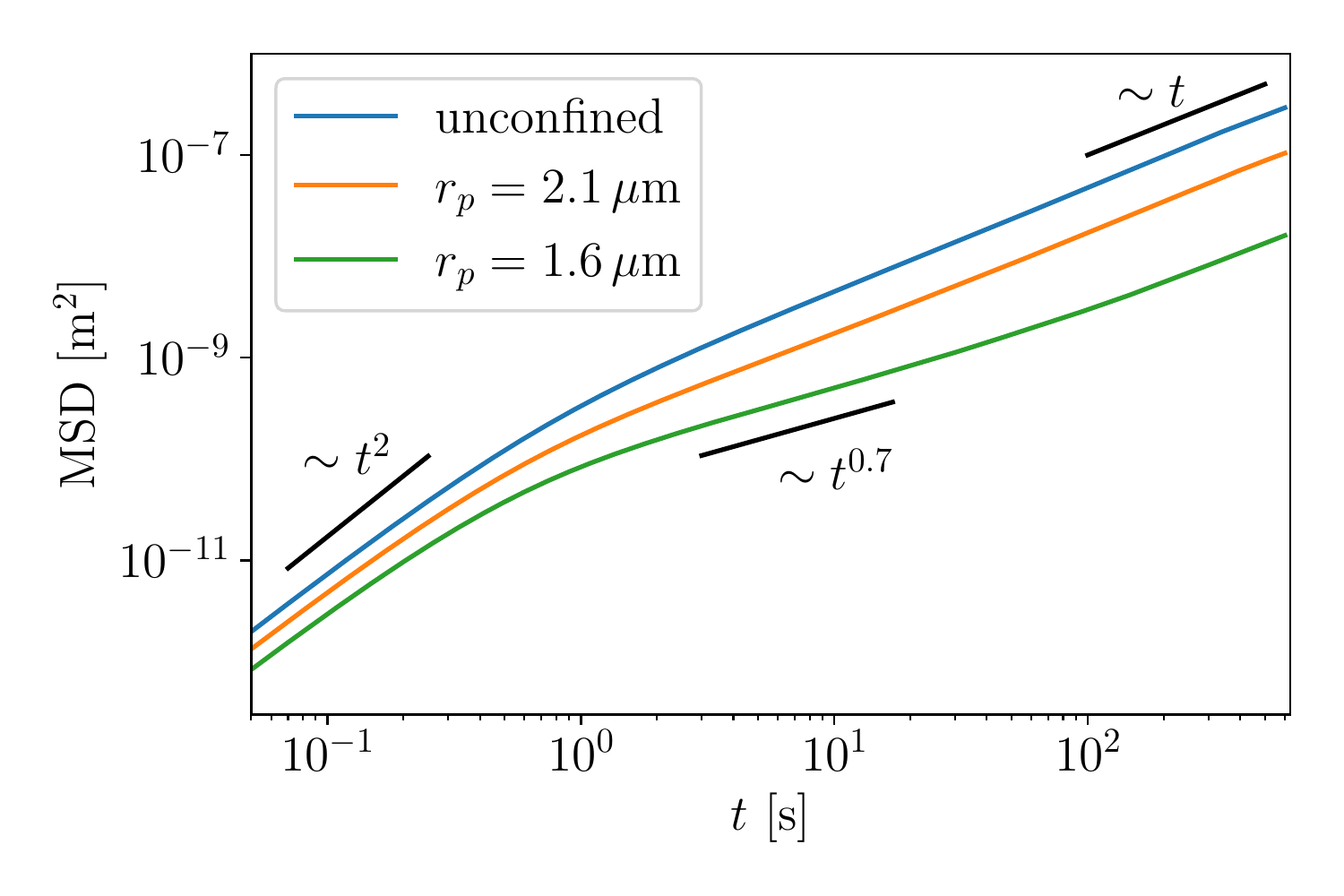}
        \caption{Mean-squared displacement of the \rnrlong{} pattern at various mean pore sizes $\meanPoreRadius$.
        The black lines indicate different scaling behaviours as a guide to the eye.}
        \label{fig:msd_rnr_subdiff}
    \end{figure}

    An example for \rnrlong{} is shown in \cref{fig:msd_rnr_subdiff}.
    It contains the qualitative features that are present in the \MSD s for all motility strategies:
    For short timescales it is super-diffusive with $\MSD(t) \sim t^2$, where the ballistic contribution of self-propulsion dominates over random motion and interactions.
    For intermediate timescales there is a sub-diffusive regime, \idEst, $\MSD(t) \sim t^\alpha$ with $\alpha<1$.
    This is a result of trapped agents that spend significant time not moving in narrow pores, waiting for a random event to allow them to escape.
    For long timescales, the motion is diffusive, \idEst, $\MSD(t) \sim 6 \diffCoeffEff t$ with an effective diffusion coefficient $\diffCoeffEff$.
    This holds true without confinement, and also in porous media as long as the confinement is not strong enough to prohibit agents from moving altogether.
    We use $\diffCoeffEff$ as the key metric to rank the different motility patterns.

\begin{figure}
        \includegraphics[width=\linewidth]{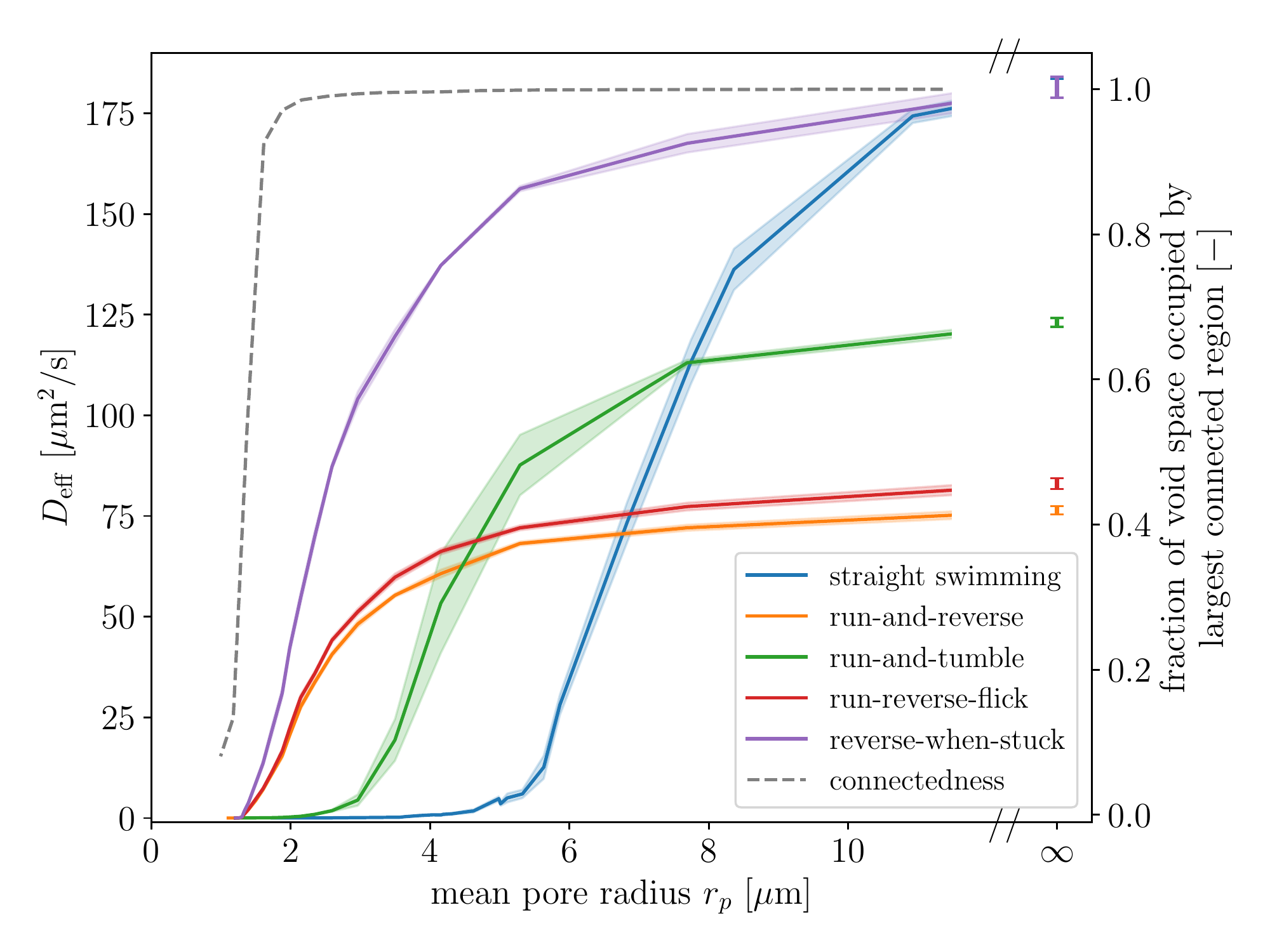}
        \caption{\revOneAdd{Solid, colored lines, left axis: }Effective diffusion coefficient \revOneAdd{$\diffCoeffEff$} as function of average pore size for all motility strategies.
        \revOneAdd{Dashed, grey line, right axis: Connectedness of the accessible void space measured as the fraction of the void space occupied by the largest connected region with local thickness $\localThickness(\vb{r}) > \bactRBead$.}
        The shaded areas denote one standard error of the mean over $\nEnsemble = 7$ statistically independent simulations.}
        \label{fig:deff_vs_pore_size}
    \end{figure}

    \Cref{fig:deff_vs_pore_size} shows $\diffCoeffEff$ as a function of mean pore radius $r_p$ of the confining geometry.
    Without confinement (mean pore radius $\meanPoreRadius \to \infty$), straight swimming leads to a larger effective diffusion than any of the other patterns, but only by a factor of about 2 to 3.
    This ratio is quite small considering that there are active reorientations in the other motility patterns while for the straight swimmers the rotational diffusion is the only source of deviation from ballistic motion.
    Due to the small size of the particles, rotational diffusion has a strong effect on swimming regardless of the specific pattern: From the rotational friction coefficient $\frictionRot$ (see Methods on how we calculate this quantity) follows the typical timescale $\rotationTimescale$ for rotational diffusion via the Einstein-Smoluchowski relation
    \begin{equation}
        \rotationTimescale = \frac{1}{2\diffCoeffRot} = \frac{\frictionRot}{2 \kbT},
    \end{equation}
    where $\diffCoeffRot$ is the rotational diffusion coefficient, $\kB$ the Boltzmann constant, and $\temperature$ the temperature.
    For the parameters of the agents simulated here (see Methods), we obtain $\rotationTimescale \approx \SI{0.7}{\second}$.
    This timescale is comparable to the typical time $\avgTRun = \SI{1}{\second}$ between active reorientations in non-straight swimming patterns, causing the relatively small ratio of $\diffCoeffEff$ between the patterns when there is no confinement.
    \revTwoAddRoundTwo{For bigger agents or flagellated bacteria, where $\rotationTimescale$ is typically much larger than $\avgTRun$, the difference in free-space diffusivity between straight swimming and the other patters would be greater.}

    In the absence of obstacles, the effective diffusion coefficient of the straight swimmer can be calculated analytically.
    It is then equivalent to the simple active Brownian particle (ABP), where the effective diffusion coefficient reads
    \begin{equation}
        \diffCoeffEff^\text{ABP} = \frac{\kbT}{\frictionTrans} + \frac{1}{3} \rotationTimescale \vSwim^2,
        \label{eq:deff_abp}
    \end{equation}
    where $\frictionTrans$ is the translational friction coefficient.
    Here, we obtain $\diffCoeffEff \approx \SI{183}{\micro\meter\squared\per\second}$ as seen in \cref{fig:deff_vs_pore_size}.

    Without confinement, the other patterns show a smaller effective diffusivity than the straight swimmers, because in addition to rotational diffusion, they use active reorientations.
    Since for \rntlong{} the average reorientation angle is $\expval{\tumbleAngle}\approx \SI{56}{\degree}$ (chosen to match the experimentally observed behaviour of \EColi~\cite{berg72a}), it results in more persistent motion than \rnrlong{} with a reorientation angle of $\SI{180}{\degree}$.
    \RRFlong{} performs slightly better than \rnrlong{} because the flicks lead to less retracing of the trajectory compared to reversals.

    For decreasing pore size, \idEst, stronger confinement, agents that employ straight swimming are the first to become ineffective at navigating through their environment.
    Even though the porous geometry is made of spheres, \idEst, convex surfaces, overlap between them can generate concave pore shapes in which elongated swimmers get stuck.
    Straight swimmers have to rely on thermal motion to randomly reorient themselves away from such pores to escape.
    Escapes are additionally hindered by the constant forward propulsion that drives them into the pore, such that translational diffusion is very unlikely to lead to a displacement out of the pore.
    The occurrence of concave, trapping pores happens at porosities where the average pore radius is still much larger than the size of the swimmer.
    Only a few of such pores significantly decrease the effective diffusivity because straight swimmers can get trapped for long durations.

\begin{figure}
        \includegraphics[width=\linewidth]{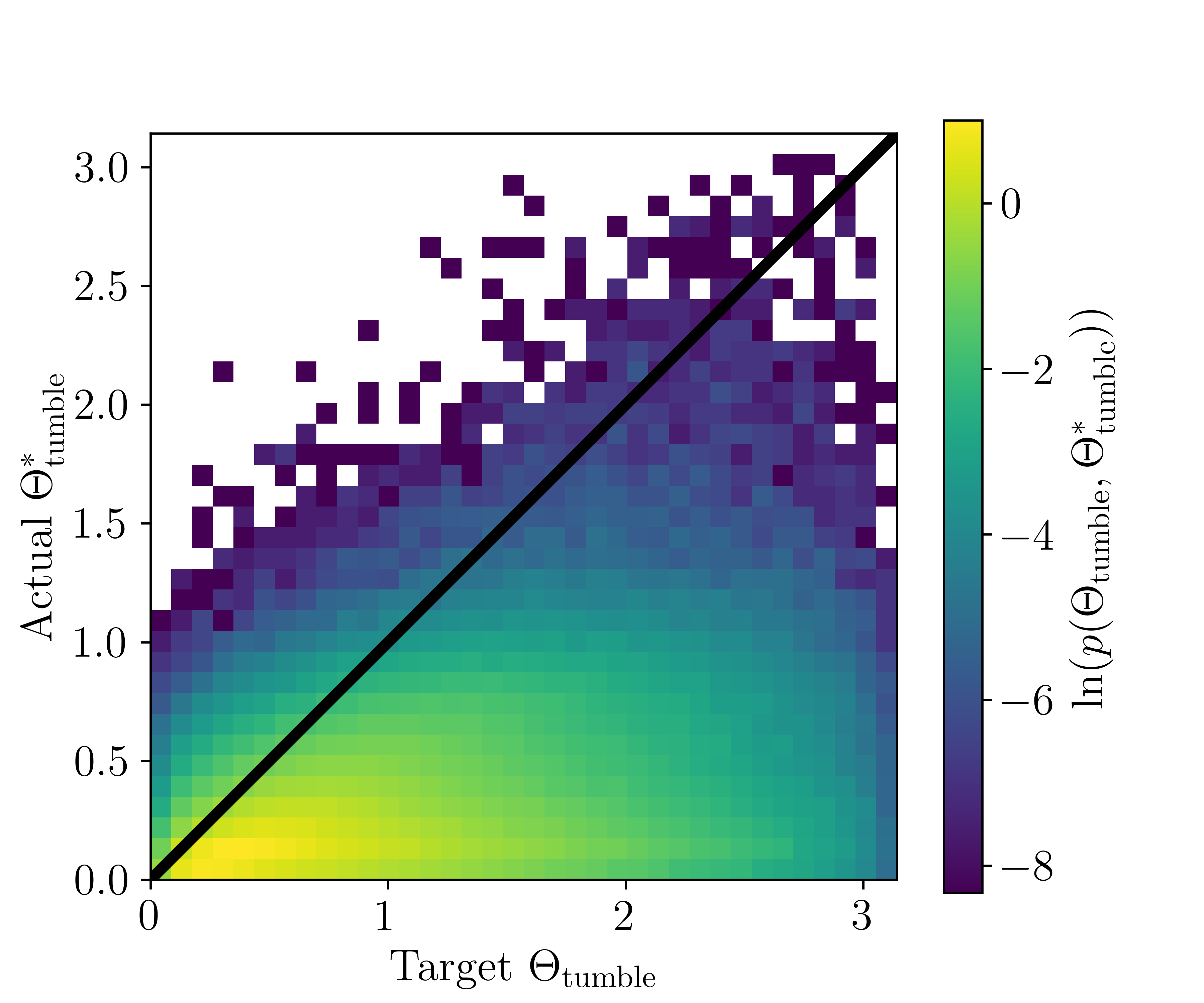}
        \caption{Joint probability density of attempted tumble angle $\tumbleAngle$ and actual tumble angle $\actualTumbleAngle$ between start and end of a tumble at mean pore radius $\meanPoreRadius \approx \SI{2}{\micro\meter}$.
        The black line indicates $\actualTumbleAngle = \tumbleAngle$.}
        \label{fig:tumble_theta_target_vs_actual}
    \end{figure}

    The next pattern to become ineffective is \rntlong, but there is a range of pore sizes where \rntlong{} outperforms straight swimming.
    Here, tumble events make it possible to escape from pores where rotational diffusion is not strong enough to lead to sufficient reorientations.
    Since the tumble angle is drawn from a distribution over $\qty[0, \pi]$, there is a probability for tumbles with $\tumbleAngle > \pi/2$, pointing the swimmer out of the pore and back to an open channel.
    Yet, the pore size at which \rntlong{} becomes ineffective is still significantly larger than that of \rrflong{} or \rnrlong.
    This is because swimmer reorientation and pore escape requires rotation of the elongated swimmer body in space, which can be suppressed by confinement.
    To illustrate this point, we show the probability density of attempted tumble angles $\tumbleAngle$ and the actual angle $\actualTumbleAngle$ between start and finish of a tumble in \cref{fig:tumble_theta_target_vs_actual} for one typical simulation.

\begin{figure}
        \includegraphics[width=\linewidth]{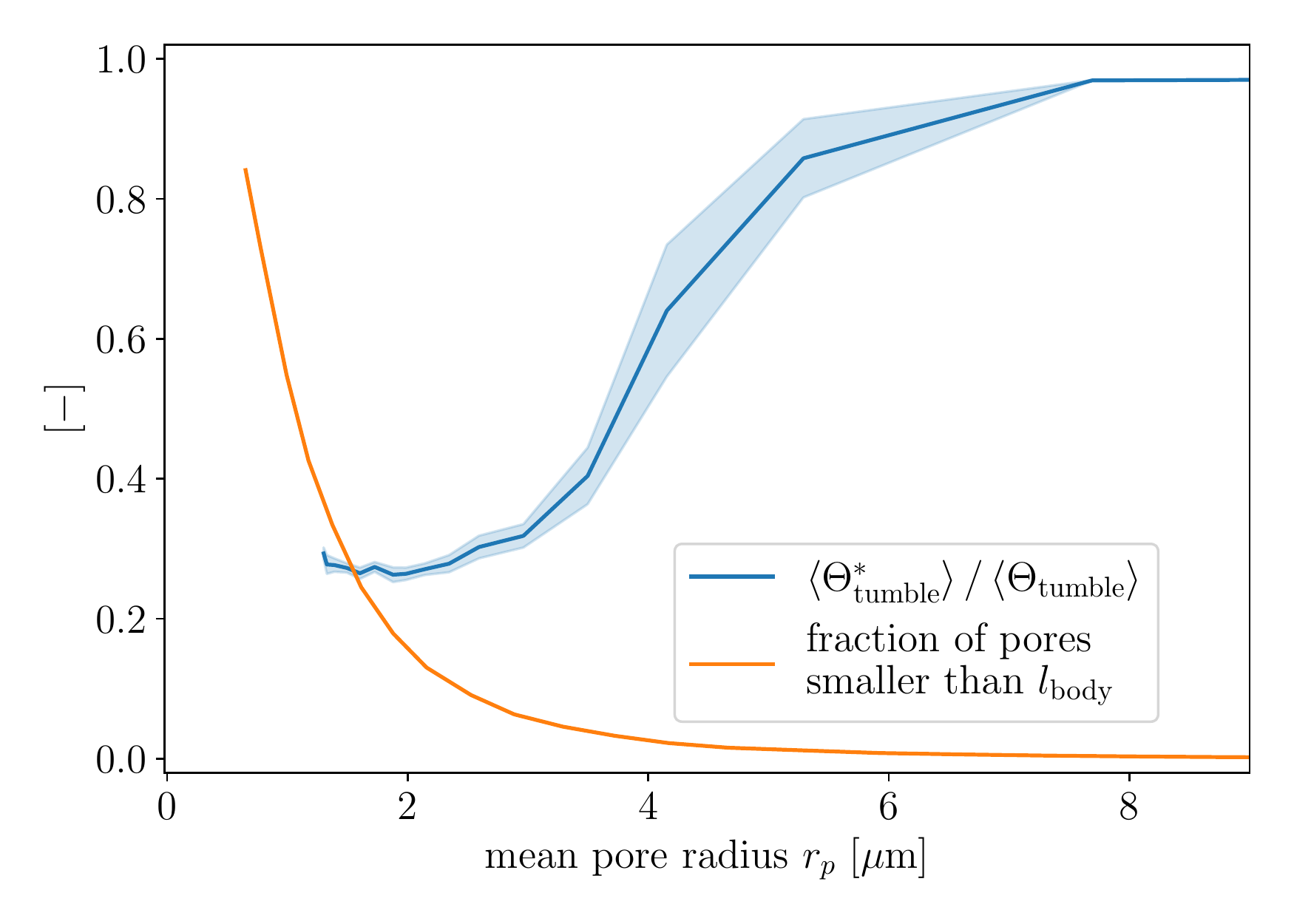}
        \caption{\revOneScratch{Three}\revOneAdd{Two} dimensionless quantities as a function of the mean pore size.
        Blue: The mean actual tumble angle $\expval{\actualTumbleAngle}$ normalized by the mean attempted tumble angle $\expval{\tumbleAngle}$.
        Orange: The fraction of pores smaller than the agent size, \idEst, with $\localThickness(\vb{r}) < \bactLBody$.
        \revOneScratch{Green: The fraction of the void space that is occupied by the largest connected region with $\localThickness(\vb{r})>\bactRBead$.}
        The shaded areas denote one standard error of the mean over $\nEnsemble = 7$ statistically independent simulations.
        For further explanations, see the main text.
        }
        \label{fig:actl_theta_vs_pore_size}
    \end{figure}

    Without rotational Brownian motion or obstacles, there would only be non-zero values on the angle bisector of the coordinate axes with magnitude according to the distribution of attempted tumble angles.
    However, in porous confinement, the deviation from $\tumbleAngle = \actualTumbleAngle$ is strongly asymmetric with the majority of actual tumble angles happening close to zero.
    Most tumbling, especially for larger angles, is suppressed by confinement, leaving agents trapped in pores despite their attempts to escape.
    To quantify this effect, we show the average actual tumble angle $\expval{\actualTumbleAngle}$ for different mean pore sizes in \cref{fig:actl_theta_vs_pore_size}.
    The suppression of tumbles with decreasing mean pore radius starts around $\meanPoreRadius \approx \SI{5}{\micro\meter}$, the same value where $\diffCoeffEff$ begins to drop significantly for \rntlong{} agents.
    We note that at this mean pore size, only a relatively small fraction of pores has a smaller radius than the length $\bactLBody$ of an agent (orange curve).
    It is enough to cause a significant deviation of $\actualTumbleAngle$ from the target tumble angle $\tumbleAngle$ because self-propelled agents are much more likely to encounter the small, trapping pores than passive particles would be:
    Active agents tend to be in contact with surfaces over long periods of time and slide along the pore walls due to their persistent motion.
    This increases the chance of entering a location of strong confinement.

    \RnRlong{} and \rrflong{} swimmers can explore environments with average pore sizes ranging down to the size of a single agent.
    This is not only because of the large probability of reversal events (certainty for \rnrlong, 50\% for \rrflong), but also because they lead to a guaranteed pore escape, unlike large tumble angles with \rntlong.
    For example, a tumble with $\tumbleAngle = \pi$ is not equivalent to a reversal event in \rnrlong.
    In the former, there needs to be enough space to allow the rotation of the swimmer body whereas in the latter, the propulsion is reversed without affecting the swimmer orientation.
    \RRFlong{} reduces to \rnrlong{} because flicks are geometrically suppressed just as tumbles are.
    Its effective diffusivity is slightly larger than that of \rnrlong{} because the smaller frequency of reversals allows the agents to move faster through open channels inbetween trapping pores.
    Both patterns become ineffective at porous media exploration only when the available void space becomes disconnected and motion is only possible within a finite volume.
    To quantify this, we calculate the volumes of connected regions with local thickness $\localThickness(\vb{r})>\bactRBead$.
    \revTwoScratch{The local thickness represents the radius of the largest sphere that contains the point $\vb{r}$ and fits entirely in the void space between the obstacles as seen in fig. 2.}
    \revOneScratch{In fig. 7 we show} \revOneAdd{The dashed grey line in \cref{fig:deff_vs_pore_size} shows} the ratio between the volume \revOneAdd{of the} largest of these regions and the total void space $V_\text{void} = \porosity L^3$, where $\porosity$ is the porosity of the porous geometry and $L$ the length of the cubic simulation domain.
    There is only one such region for $\meanPoreRadius \gtrsim \SI{2}{\micro\meter}$, but around $\meanPoreRadius \approx \SI{1.5}{\micro\meter}$ the void space splits into many smaller regions such that even for the larger ones there can be no more percolating motion through the simulation box.

    Kurzthaler \textit{et al.}~\cite{kurzthaler21b} find that there is no significant difference between the effective diffusivity of \rntlong{} and \rnrlong{} in porous media.
    However, their implementation of \rntlong{} includes a $50\%$ chance of reversing when tumbling, so we would classify that pattern as run-and-tumble-or-reverse.
    According to our observation of suppressed tumbles, this pattern will reduce to \rnrlong{} when sufficiently confined, at which point our results are in agreement with theirs.

    \RnRlong{} and \rrflong{} are the best biologically inspired patterns for porous media exploration at very small pore sizes, but they do not perform well for larger porosities, where straight swimming is optimal.
    This inspired the creation of the \rwnlong{} pattern, combining the best features of straight swimming and \rnrlong, especially propulsion reversal without rotation of the swimmer body.
    As expected, it results in the largest effective diffusivity and therefore best exploration efficiency over the whole range of pore sizes.
    At very small pore sizes, \rwnlong{} performs better than \rnrlong{} and \rrflong, because the agent only performs reversals when they are needed for pore escape.
    When it has found an open channel through the porous medium, it follows that channel until it gets stuck at the end without being interrupted by a randomly triggered reversal event.

    \revOneAdd{
    \subsection{Run time variation}
    While the previous section focuses on the influence of confinement on the dynamics of agents with fixed parameters, we now explore the influence of motility strategy parameters with fixed level of confinement.
    We assume the agents' shape and propulsion speed to be given and vary only the internal parameters that are directly linked to the strategy.
    We choose the run time parameter $\avgTRun$ for comparison as it is common amongst the biologically inspired patterns \rnrlong, \rntlong{} and \rrflong{} as shown in \cref{fig:trun_var}.
    We do not change tumble or flip parameters, because they interpolate between straight swimming and \rntlong{} as well as between \rnrlong{} and \rrflong{}, and we want to preserve the essential features of each motility strategy.
    \begin{figure}
    \includegraphics[width=\linewidth]{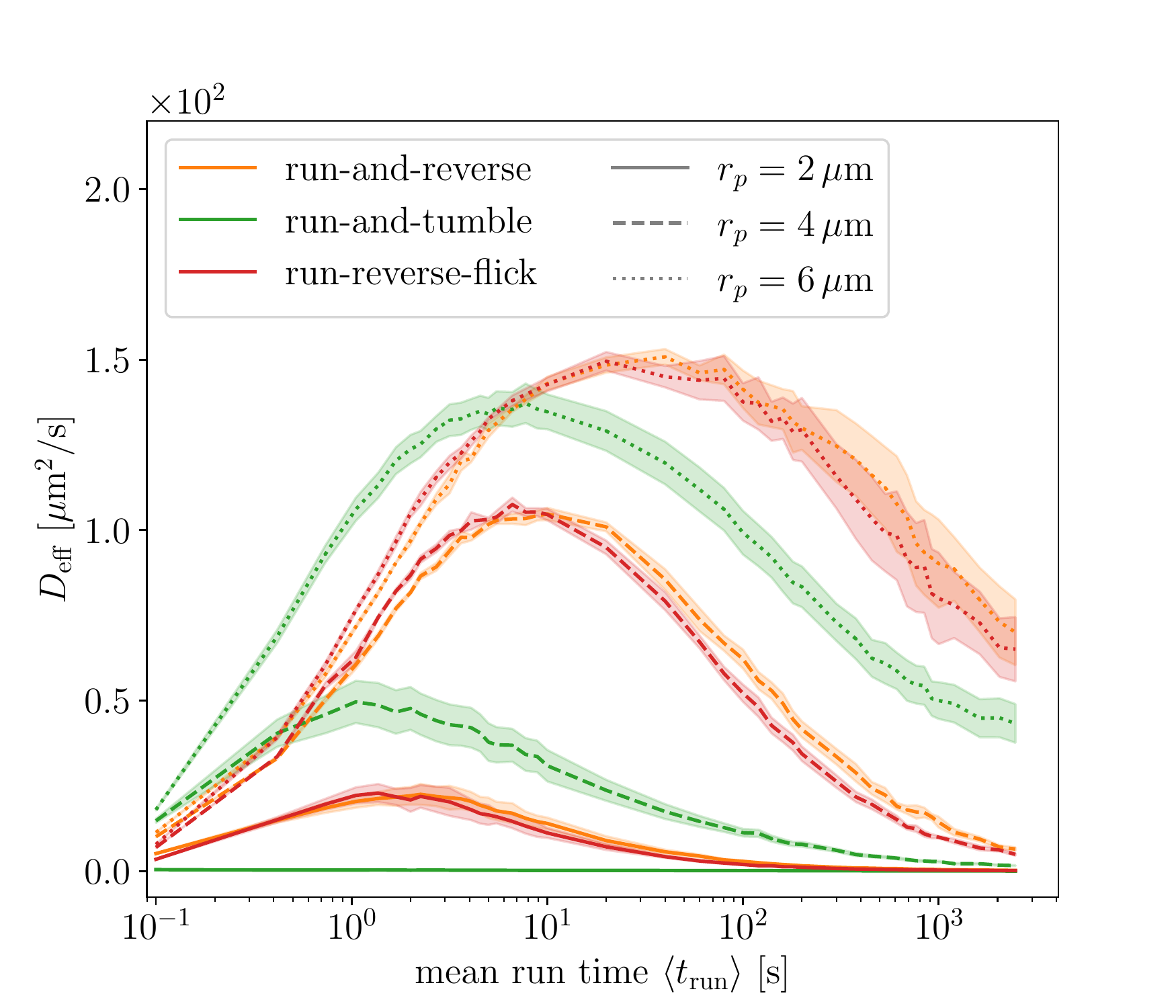}
    \caption{\revOneAdd{Effective diffusion coefficient as a function of mean run time.
    Colors denote the motility stategy.
    Solid, dashed and dotted lines correspond to different mean pore sizes.
     The shaded areas denote one standard error of the mean over $\nEnsemble= 5$ statistically independent simulations.}}
     \label{fig:trun_var}
    \end{figure}
    Simulations are performed at three levels of confinement $\meanPoreRadius \in \qty{\SI{2}{\micro\meter}, \SI{4}{\micro\meter}, \SI{6}{\micro\meter}}$ where we expect to see the most geometry-related differences between the motility strategies.}

\revOneAdd{
    All motility strategies perform very badly with $\avgTRun \to 0$, as in this limit, agents reduce to particles without active propulsion that rely solely on thermal diffusion for exploration of their environment.
}

\revOneAdd{
    For the largest mean pore radius $\meanPoreRadius = \SI{6}{\micro\meter}$,
    \revOneScratchRoundTwo{the effective diffusion coefficient grows monotonically before it reaches a plateau at the same value for all strategies.
    Here, trapping events are rare and both reversal and tumbling leads to a pore escape. For small mean run times, \rntlong{} is advantageous because the active reorientation is smaller than for the other patterns.
    For large mean run times $\avgTRun \gg \rotationTimescale$, active reorientation has a negligible effect compared to Brownian rotation, such that the motility pattern becomes irrelevant as long as it can facilitate an escape from the very few trapping pores.
    }
    \revOneAddRoundTwo{\rntlong{} is most efficient for short mean run times $\avgTRun \lessapprox \SI{7}{\second}$.
     In this regime, agents perform active reorientations frequently, so they spend more time swimming through open channels in the porous medium than being trapped.
     Here, \rntlong{} agents benefit from larger persistence than the patterns with reversals.
     However, tumbles are less likely to lead to an escape from narrow pores than reversals.
     This is why with increasing $\avgTRun$, the effective diffusivity for \rntlong{} reaches its maximum earlier and at a lower value than the effective diffusivity of the other two patterns.
     For long mean run times, swimmers spend more time trapped than swimming freely due to less frequent escape attempts.
     Pore escape efficiency becomes more important than large persistence in open channels, so \rnrlong{} and \rrflong{} perform best.
     In the limit $\avgTRun\to\infty$, all motility patterns reduce to straight swimming and must have the same $\diffCoeffEff$.
     In our simulations, this limit will be reached only at mean run times $\avgTRun \gg \SI{2000}{\second}$, far beyond biologically reasonable timescales.
     }
}

\revOneAdd{\revOneScratchRoundTwo{
    At intermediate confinement, \rnrlong{} and \rrflong{} are still able to escape trapping pores qickly such that they benefit from elongated run phases and $\diffCoeffEff$ grows monotonically.
    \RnTlong{} agents can still escape trapping pores in general, but at $\meanPoreRadius = \SI{4}{\micro\meter}$, tumbles start to be suppressed (see fig. 7), such that more attempts are needed to escape a pore.
    Therefore, \rntlong{} shows nonmonotonic behaviour: After an optimum of $\avgTRun \approx \SI{1}{\second}$, the benefit of longer run periods does not outweigh the cost of longer trapping durations and $\diffCoeffEff$ decreases.}}

\revOneAdd{
\revOneScratchRoundTwo{
    In the case of strong confinement, \rntlong{} agents are not able to explore the medium at any mean run time, as tumbling is completely suppressed.
    \RnRlong{} and \rrflong{} agents show nonmonotonic behaviour as observed and explained in Ref.~\cite{kurzthaler21b}: When the length of free paths in the porous medium coincides with the average run length $l_\text{run} = \avgTRun \vSwim$, the effective diffusivity is optimal.
    Deviations lead to a decrease in $\diffCoeffEff$, either due to premature cancellation of a run phase that could have followed a free path for a longer time or due to long trapping times that could have been avoided by more frequent reversals.
    }
}

\revOneAddRoundTwo{
    For intermediate and tight confinement ($\meanPoreRadius = \SI{4}{\micro\meter}, \SI{2}{\micro\meter}$), the same qualitative behaviour is observed.
    However, because open channels are fewer and shorter, and trapping pores are encountered more often, pore escape by motility reversal becomes more relevant at smaller $\avgTRun$.
    At $\meanPoreRadius = \SI{4}{\micro\meter}$, \rntlong{} shows only a marginally larger effective diffusivity than \rnrlong{} and \rrflong{} for $\avgTRun \lesssim \SI{0.7}{\second}$.
    For larger mean run times, the performance is significantly worse.
    This mean pore size coincides with the $\meanPoreRadius$ at which tumbles are still possible, but begin to be suppressed, see \cref{fig:actl_theta_vs_pore_size}, reducing the efficiency of pore escape.
    At $\meanPoreRadius = \SI{2}{\micro\meter}$, \rntlong{} agents are not able to explore the medium at any mean run time, as tumbling is completely suppressed.
    }

\revOneAddRoundTwo{
    The nonmonotonic behaviour and shift in the maximum of $\diffCoeffEff$ for all motility patterns was observed and explained for \rnrlong{} in Ref.~\cite{kurzthaler21b}: When the length of free paths in the porous medium coincides with the average run length $l_\text{run} = \avgTRun \vSwim$, the effective diffusivity is optimal.
    Deviations lead to a decrease in $\diffCoeffEff$, either due to premature cancellation of a run phase that could have followed a free path for a longer time or due to long trapping times that could have been avoided by more frequent reversals.
}

\revOneAdd{
    In general, these observations support the qualitative conclusions drawn from the simulations performed at fixed $\avgTRun$.
    For all three levels of confinement, the ranking of motility strategies remains the same \revOneScratchRoundTwo{regardless of $\avgTRun$}\revTwoAddRoundTwo{for small $\avgTRun$}.
    However, by tuning the run length, \rnrlong{} agents can \revOneScratchRoundTwo{approach or} outperform \rntlong{} agents.
    We therefore suggest that if manufacturing or genetically creating a \rwnlong{} agent is not feasible, optimizing a \rnrlong{} swimmer is the best choice when creating an agent for porous media exploration.
}

    \section{Conclusion}
    \label{sec:conclusion}

    We have performed Langevin dynamics simulations of rod-shaped, self-propelled and self-steered agents with various motility patterns in porous model geometries spanning a large range of porosities and pore sizes.
    By quantifying their long-time, effective diffusivity, we evaluated their ability to explore these porous environments: At high porosity, \idEst{}, large pore sizes, straight swimming performs best due to the absence of active rotation, hence the trajectories are ballistic for the longest possible time.
    At intermediate pore sizes, \rntlong{} has the largest effective diffusivity.
    Here, the agents can escape the pores by tumbling, which straight swimmers cannot, and they can explore the larger pore spaces with a more persistent motion than \rnrlong{} or \rrflong{}.
    At very small pore sizes, rotation of the rods is suppressed by confinement, causing \rntlong{} swimmers to get stuck and preventing \rrflong{} swimmers from flicking.
    In this regime, \rnrlong{} and \rrflong{} swimmers can still escape small pores because their reversal mechanism enables them to reverse propulsion without rotation of the agent itself, making them the \revOneScratch{optimal biologically inspired motility patterns we considered here.} \revOneAdd{only viable strategies in tight confinement.}
    \revOneAdd{
    By optimizing the run length of the different motility strategies and thus exhausting the full potential of each pattern, we find that in all geometries investigated, \rnrlong{} can \revOneScratchRoundTwo{approach or} surpass the other strategies, making it the optimal biologically inspired motility pattern considered in this study.
    }

    These results prompted us to develop a motility pattern that outperforms the biologically inspired patterns by endowing the active agents with memory and the ability to sense position (or velocity) for some time span, and an intelligence feature that makes a decisions based on this memory:
    If the agent only reverses when it is stuck, defined as not moving more than its own length in its memory time, it can optimally explore open channels in the porous geometry while still being able to escape trapping pores.

    With or without intelligence, we suggest that being able to reverse propulsion without rotation of the agent itself should be a high priority when designing active micro-agents for medical and engineering applications in confined spaces.
    Only with this ability they can efficiently navigate the inevitably porous geometries in which they are deployed.
    After all, the need for miniaturisation of agents in these applications arises from the highly confined environments in which their tasks are to be performed.
    Furthermore, our results can serve as a basis for developing other optimized navigation strategies for specific environments.

    \section{Methods}
    \label{sec:model}

    \subsection{Particle model}
    The Langevin equations of motion for the particle positions $\partclPos_i$ in three dimensions read

    \begin{equation}
        m \ddot{\partclPos}_i = -\frictionTrans \dot{\partclPos}_i + \frac{\vSwim}{\frictionTrans} \partclDirec_i + \vb{F}_i(\partclPos_i, \partclDirec_i) + \sqrt{2 \frictionTrans \kbT} \noiseTrans_i.
        \label{eq:langevin_trans}
    \end{equation}

    Here, $m$ is the particle mass, $\partclDirec$ a unit vector describing the particle orientation, $\vb{F}_i$ an external force from particle-boundary interactions, and $\noiseTrans(t)$ a random noise vector with $\expval{\noiseTrans} = \vb{0}$ and $\expval{\noiseTrans_i(t)\noiseTrans_j(t')} =  \delta_{ij}\delta(t-t')\vb{1}$, where $\expval{\cdot}$ denotes an ensemble average and $\vb{1}$ the identity matrix in three dimensions.
    For the particle orientations $\partclDirec_i$ we have analogously

    \begin{align}
        \outerdot{\partclDirec}_i = \partclAngVel_i \cross \partclDirec_i, \\
        I \dot{\partclAngVel_i} = -\frictionRot \partclAngVel_i + \frac{\activeAngVel}{\frictionRot} \orthoVec_i + \vb{M}_i(\partclPos_i, \partclDirec_i) + \sqrt {2 \frictionRot \kbT} \noiseRot_i,
        \label{eq:langevin_rot}
    \end{align}

    where $I$ is the particle moment of inertia tensor, $\partclAngVel_i$ the angular velocity vector, \revTwoAdd{$\activeAngVel$ the active angular velocity analogous to $\vSwim$}, $\orthoVec_i$ a unit vector perpendicular to $\partclDirec_i$ \revTwoAdd{selected at random by the motility pattern}, $\vb{M}_i$ an external torque stemming from interactions, and $\noiseRot$ a noise term with the same properties as $\noiseTrans$.
    All simulations are performed using \ESPResSo~\cite{weik19a} to integrate the equations of motion.

    Our active agents are constructed from $\bactNBeads = 5$ \revTwoScratch{point} particles that are rigidly connected in a rod-like manner as shown in \cref{fig:model}.
    Only the position and orientation of the central particle \revTwoAdd{at $\partclPos_i$} are propagated in time using \cref{eq:langevin_trans} to \cref{eq:langevin_rot}.
    All other particles \revTwoAdd{associated with the agent} \revTwoScratch{transfer their forces and torques to the central particle and} are repositioned \revTwoAdd{according to their constant configuration relative to the central particle after every time step} \revTwoScratch{after propagating the central particle} such that the rod behaves like a rigid body.
    \revTwoAdd{The force and torque on the central particle are calculated from
    \begin{equation}
        \vb{F}_i = \sum_{j=1}^{\bactNBeads} \vb{F}\qty(\partclPos_i^{(j)}),
        \label{eq:rod_force}
    \end{equation}
    \begin{equation}
        \vb{M}_i = \sum_{j=1}^{\bactNBeads} (\partclPos_i^{(j)} - \partclPos_i) \cross \vb{F}\qty(\partclPos_i^{(j)}),
        \label{eq:rod_torque}
    \end{equation}
    where $\partclPos_i^{(j)}$ is the position of the $j^{th}$ particle of agent $i$ and $\vb{F}$ the force derived from an interaction potential detailled below.}

    In our simulations we do not consider interactions between agents as we want to analyse only single agent properties.
    The mass $m$ and moment of inertia $I$ are calculated by approximating the rods as cylinders with constant density $\rho$.
    However, we show later that the exact values of $m$ and $I$ do not alter the physical behaviour of the agents.

    To obtain the translational friction coefficient $\frictionTrans$, we approximate the rods as a spheroids and use the results of Datta \& Srivastava~\cite{datta99a}.
    Taking half of the rod length $\bactLBody$ and the radius $\bactRBead$ as the long and short half-axis, respectively, the friction coefficient is calculated \via{}

    \begin{align}
        e = \sqrt{1 - \qty(\frac{\bactRBead}{\bactLBody/2}) ^ 2}, \nonumber \\
        \frictionTrans = \frac{16  \pi    e ^ 3\bactLBody/2}{(1 + e ^ 2)  \ln[(1 + e) / (1 - e)] - 2 e} \dynVisc,
    \end{align}

    where $\dynVisc$ is the dynamic viscosity of the surrounding medium.

    Rotational Brownian motion has a strong influence on the dynamics of self-propelled particles as it sets the persistence of active, ballistic motion.
    It is therefore vital to obtain a good estimate of the rotational friction coefficient $\frictionRot$ of our agents.
    For rotations around equatorial axes (\idEst, axes perpendicular to the symmetry axis) it is calculated from Perrin theory~\cite{perrin34a} \via

    \begin{align}
        p = \frac{\bactLBody/2}{\bactRBead}, \quad \xi = \frac{\sqrt{p ^ 2 - 1}}{p}, \nonumber \\
        F_{eq} = \frac{2}{3}  \frac{p ^{-2} - p ^ 2}{1 - (2 - p ^{-2})\text{atanh}\qty(\xi) / \xi}, \nonumber \\
        \frictionRot = F_{eq}  8 \pi \dynVisc \frac{\bactLBody }{2} \bactRBead^ 2.
    \end{align}

    Rotations around the axis of symmetry are neglected as they do not affect any observable of the system.

    The self-propulsion and self-rotation that separates our model of active agents from passive colloids is determined by $\vSwim$ and $\activeAngVel$.
    According to the specific motility pattern, these terms can be constant or time dependent.
    The motility pattern is evaluated and the active forces and torques are updated with a time step $\timeSlice$.
    It is an integer multiple of the simulation timestep $\timeStep$ used in the velocity-Verlet scheme to integrate the equations of motion.
    This reflects the difference in time scales between the Brownian motion and changes in motility, and speeds up simulations significantly.

    \subsection{Porous media model}
    \label{subsec:porous-media-model}
    Inspired by the experimental setup of Bhattacharjee \& Datta~\cite{bhattacharjee19a}, we model the porous environment with spheres of radius $\radiusHydroSphere$.
    The spheres are placed randomly throughout the simulation box and fixed in space for the entire duration of the simulation.
    As an approximation of hardcore repulsion, all individual particles of the swimmer rods interact with all spheres with a truncated and shifted purely repulsive Lennard-Jones potential

    \begin{align}
        V(r) = &4 \epsilon \left[ \left(\frac{\sigma}{r-\radiusHydroSphere} \right)^{12} - \left(\frac{\sigma}{r-\radiusHydroSphere} \right)^{6} + \epsilon   \right] \nonumber \\
        &\cross H(\radiusHydroSphere + \bactRBead - r),
    \end{align}

    where $r$ is the distance between the particle and the sphere center, $\epsilon$ the interaction strength, $\sigma = 2^{-\frac{1}{6}}\bactRBead$ and $H(\cdot)$ the Heaviside step function.
    All simulations are performed in a cubic, $L\cross L \cross L$ domain with periodic boundary conditions, where $L$ denotes the simulation box size.
    The control parameter for the porous geometry is the number of spheres.
    To analyse the porous geometry, we first use the positions of the spherical obstacles to generate a binary image of the pore space at a resolution of $\Delta x = \SI{0.25}{\micro\meter}$.
    Then we use the \porespy~\cite{gostick19a} python package to obtain quantitative measures such as
    porosity $\porosity$, local thickness $\localThickness(\vb{r})$, and the pore size distribution.
    \revTwoAdd{The local thickness represents the radius of the largest sphere that contains the point $\vb{r}$ and fits entirely in the void space between the obstacles as seen in \cref{fig:porous_slice_local_thickness}.
    The mean pore radius $\meanPoreRadius$ is determined by calculating the mean of $\localThickness(\vb{r})$ in the regions where it is non-zero.}

    \subsection{Parameter choice}
    \label{subsec:parameter-choice}
    To compare the motility patterns against each other, we choose the same physical parameters for all simulations: $\bactLBody = \SI{2}{\micro\meter}$, $\bactRBead = \SI[parse-numbers = false]{1/3}{\micro\meter}$, $\vSwim = \SI{28}{\micro\meter\per\second}$, $T = \SI{300}{\kelvin}$, $\dynVisc = \SI{8.9e-4}{\pascal\second}$, $\radiusHydroSphere = \SI{5}{\micro\meter}$, $\epsilon = \kbT$ and $L = \SI{80}{\micro\meter}$.
    \revOneAdd{In Ref.~\cite{darus} we report effective diffusivities for $\radiusHydroSphere = \SI{10}{\micro\meter}$.
    The results are in very good agreement with \cref{fig:deff_vs_pore_size}, hinting at the independence of our conclusions from the details of the porous medium model.}
    \revOneAdd{Unless noted otherwise, we} \revOneScratch{We} set the average run times for \rnrlong, \rntlong{} and \rrflong{} to $\expval{t_{run}}= \SI{1}{\second}$, the average time of rotation for \rntlong{} and \rrflong{} to $\avgTTumble = \tFlick = \SI{0.1}{\second}$ and the memory time for \rwnlong{} to $\tMemory = \SI{1}{\second}$.
    For \rntlong, we set $\tumbleDRot = \SI{5}{\per\second}$, which results in $\expval{\tumbleAngle} \approx \SI{56}{\degree}$, close to values observed in \EColi~\cite{berg72a}.
    These might not be the optimal parameters for each of the patterns for all pore sizes, but they serve as a common ground for the evaluation of the pattern performance.

    For agents of this size at the density of water $\rho_\text{water} = \SI{1e3}{\kilo \gram\per\meter\cubed}$, the diffusive relaxation time $\tau_\text{relax}= m/\frictionTrans \approx \SI{7e-8}{\second}$ is very small compared to all other timescales of the system, so the dynamics is overdamped and the exact value of $m$ does not influence the physical behaviour.
    In our simulations, we set the density of the agents to $\rho = 10^5 \rho_\text{water}$.
    \revTwoScratchRoundTwo{This still leaves the dynamics overdamped but allows us to choose larger time steps of $\timeSlice=15\timeStep=\SI{5e-3}{\second}$.}
    \revTwoAddRoundTwo{We therefore do not resolve the actual, negligibly fast, momentum relaxation.
    This allows us to choose large time steps $\timeSlice=15\timeStep=\SI{5e-3}{\second}$ and thereby speed up the simulations by five orders of magnitude.}

    Simulations are performed with $\nBacts = 100$ agents and run for $\simDuration = \SI{6000}{\second}$ to collect a sufficient amount of stochastic data, with an additional $\SI{600}{\second}$ warm-up phase before data collection starts.
    They are repeated \revOneScratch{at least} $\nEnsemble \revOneScratch{= 7}$ times with different random seeds, \idEst, different geometries, particle starting positions and noise realisations.
    Error quantifications shown in the previous sections represent the standard error of the mean over different simulations.

    \section*{Code and data availability}
    The data that support the findings of this
    study as well as the source code are available \revOneScratch{from the corresponding author upon reasonable request} \revOneAdd{at Ref.~\cite{darus}}.

    \section*{Acknowledgments}
    Funded by the Deutsche Forschungsgemeinschaft (DFG, German
    Research Foundation) - under Project Number 327154368-SFB 1313 and under Germany's Excellence Strategy - EXC 2075 - 390740016.
    We thank Sujit S. Datta for inspiring discussions.

    \section*{Author contributions}
    C.H. designed the overall study.
    C.L. developed the numerical model, performed the simulations and analysed the data.
    C.L. and C.H. wrote the manuscript.

    \section*{Competing interests}
    The authors declare no competing interests


\end{document}